\documentclass[10pt,letterpaper]{article}
\usepackage[top=0.85in,left=2.75in,footskip=0.75in]{geometry}
\usepackage{amsmath,amssymb}
\usepackage{changepage}					
\usepackage{fixltx2e}
\usepackage[utf8x]{inputenc} 				
\usepackage{textcomp,marvosym}		
\usepackage{cite}									
\usepackage{nameref,hyperref}			
\usepackage[right]{lineno}					
\usepackage{microtype}						
\DisableLigatures[f]{encoding = *, family = * }
\usepackage[table]{xcolor}					
\usepackage{array}								
\newcolumntype{+}{!{\vrule width 2pt}}
\newlength\savedwidth							

\newcommand\thickhline{\noalign{\global\savedwidth\arrayrulewidth\global\arrayrulewidth 2pt}%
\hline
\noalign{\global\arrayrulewidth\savedwidth}}
\usepackage{setspace} 
\usepackage{ccaption}
\usepackage{graphicx}
\graphicspath{ {images1/} }

\raggedright
\usepackage[aboveskip=1pt,labelfont=bf,labelsep=period,justification=raggedright,singlelinecheck=off]{caption}

\makeatletter
\renewcommand{\@biblabel}[1]{\quad#1.}
\makeatother

\usepackage{lastpage,fancyhdr,graphicx}
\usepackage{epstopdf}
\fancyheadoffset[L]{2.25in}
\fancyfootoffset[L]{2.25in}
\begin{document}
\vspace*{0.2in}

\begin{flushleft}
{\Large
\textbf\newline{ Proxyeconomics, the inevitable corruption of proxy-based competition} 
}
\newline
\\
Oliver Braganza\textsuperscript{1,*},
\\
\bigskip
\textbf{1} Institute for Experimental Epileptology and Cognition Research, University of Bonn, Germany
\bigskip

* oliver.braganza@ukbonn.de

\end{flushleft}
\section*{Abstract}
When society maintains a competitive system to promote an abstract goal, competition by necessity relies on imperfect proxy measures. For instance profit is used to measure value to consumers, patient volumes to measure hospital performance, or the Journal Impact Factor to measure scientific value. Here we note that \textit{any proxy measure in a competitive societal system becomes a target for the competitors, promoting corruption of the measure}. This suggests a general applicability of what is best known as Campbell's or Goodhart's Law. Indeed, prominent voices have argued that the scientific reproducibility crisis, the financial crisis of 2008, and inaction to the threat of global warming represent instances of such competition-induced corruption. Moreover, competing individuals often report that competitive pressures limit their ability to act according to the societal goal. This suggests some kind of \textit{lock-in}, i.e. a powerful mechanism locking the system into a certain subset of possible outcomes. However, despite the profound implications, we lack a coherent theory of such a process, potentially due to problems posed by traditional disciplinary boundaries. Here we propose such a theory, formalized as an agent-based model, integrating insights from complex systems theory, contest theory, behavioral economics and cultural evolution theory. The model reproduces empirically observed patterns and makes predictions at multiple levels. It further suggests that any system is likely to converge towards an equilibrium level of corruption determined by i) the information captured in the proxy and ii) the strength of an intrinsic incentive towards the societal goal. Overall, the theory offers mechanistic insight to subjects as diverse as the scientific reproducibility crisis and the lack of an appropriate response to global warming.



\section{Competitive societal systems}
In complex societal systems, any competition to promote an abstract goal must by necessity rely on proxy measures to rank agents with respect to their performance. This suggests the applicability of Campbell's or Goodhart's Law to any proxy-based competition \cite{Campbell1979,Goodhart1984, Strathern1997}:  \textit{Any proxy measure in a competitive societal system becomes a target for the competing individuals (or groups), promoting corruption of the measure}. In other words, societal competition should be generally expected to create an informational difference between proxy measures and the societal goals they were designed to promote. Importantly, this should happen both by active behaviors (gaming) but also by a range of purely statistical mechanisms \cite{Manheim2018A, Smaldino2016}.

The societal system performs the dual task of collecting the information to create the proxy measure(s) and creating the institutions that implement competition. While modern societies increasingly rely on competitive systems and the proxy measures they require, there are prominent arguments and compelling evidence suggesting corruption of the measures (Table \ref{tab:tab1} \cite{Nichols2005, Koretz2008, Wilsdon2015, Fire2018, Benabou2016, Baker2016, Fochler2016, Smaldino2016, Stiglitz2010, Jackson2009, Gigerenzer2013, Gross2019}).  Indeed, it is conceivable that a highly proxy-oriented system leads to substantial effort- and resource misallocation with detrimental effects on the actual societal goal. Importantly, this would remain hidden as long as the proxy persisted as the central evaluative tool of system performance. We suggest the term proxy-economy for a highly proxy-oriented, or in Campbell's terms ``corrupted'', system. In such a system a large portion of activities and resources would be devoted toward the proxy measures without furthering the actual societal goal. For instance in academia, a substantial fraction of non-reproducible research across scientific disciplines, as well as subjective researcher assesments, suggest an excessive proxy orientation \cite{Baker2016}. Similarly, a strong case can be made that rich economies as a whole have become mostly proxy-oriented \cite{Kubiszewski2013, Wilkinson2009, Stiglitz2010, Jackson2009, vandenBergh2017}. Consider for instance, that subjective as well as objective measures of well-being are logarithmically related to income (GDP) \cite{Stevenson2008, Kahneman2010}. In other words, the relevant returns (societal goal) to economic growth (proxy) are exponentially diminishing. Indeed, among rich nations the beneficial effect of economic growth is dwarfed by other sources of variability, and quite challenging to even detect statistically \cite{Wilkinson2009, Stevenson2008}. At the same time, the material footprint of economic growth, including environmental externalities such as CO\textsubscript{2} emissions, are linearly related to income (GDP), implying that we are likely substantially underestimating negative welfare consequences of economic growth.

Beside such macro-level arguments, another rich source of supporting evidence results from critical examination of individual practices under competing accounts of proxy- or goal-orientation provides  \cite{Smaldino2016, Benabou2016, Gabaix2005, Braganza2019}. For instance, it has recently been observed, that scientific sample size choices can be readily explained by a competitive economic (proxy-driven) account, but are difficult to reconcile with scientific deliberations (goal-driven)\cite{Braganza2019}. Finally, it is important to point out that the mentioned examples appear to display what might be called a \textit{lock-in} effect, i.e. persistent corruption despite widespread acknowledgement of the arising problems, combined with accounts of limited agency. Understanding the nature and mechanism of such \textit{lock-in} is likely to prove crucial to addressing phenomena ranging from the scientific reproducibility crisis to combating anthropogenic climate change.
\\
\begin{table}[!ht]
\begin{adjustwidth}{-2.25in}{0in} 
\centering
\caption{{\bf Illustratory examples of goals, proxies, and corruption claims in proxy-based competititive systems.}}
\begin{tabular}{|>{\bfseries}p{1.8cm}||p{2cm}|p{2cm}|p{2cm}|p{2cm}|p{2cm}|}
\hline
& \textbf{Science}   & \textbf{Medicine}  & \textbf{Education}& \textbf{Politics}   & \multicolumn{1}{c}{\textbf{Markets}} \\ \thickhline
\textcolor{blue}{Societal Goal} & \textcolor{blue}{true and relevant research} & \textcolor{blue}{patient health} & \textcolor{blue}{knowledge, skills} & \textcolor{blue}{voter  representation} & \textcolor{blue}{subjective and objective well-being}\\ \hline
\textcolor{red}{Proxy Measure} & \textcolor{red}{publication count, impact factor} & \textcolor{red}{patient numbers, profit}& \textcolor{red}{standardized test scores} & \textcolor{red}{publicity, votes} & \textcolor{red}{income, profit, GDP}\\ \hline
Corruption Claims & reproducibility crisis  \cite{Wilsdon2015, Fire2018, Fochler2016, Smaldino2016, Gross2019}& bad patient care, overtreatment \cite{Gigerenzer2013, Poku2016} & teaching to the test \cite{Nichols2005, Koretz2008} & populism, lobbycracy \cite{Gowdy2016, Pluchino2011} & financial crises, global warming\cite{Benabou2016, Stiglitz2010, Jackson2009}  \\\hline
\end{tabular}
\begin{flushleft} Any competitive societal system to achieve and abstract goal must rely on proxy measures. However, and proxy measure becomes a target for the competing individuals (or groups).  Campbell's and Goodhart's Laws state that this will promote corruption of the measures. \cite{Goodhart1984, Campbell1979, Strathern1997}
\end{flushleft}
\label{tab:tab1}
\end{adjustwidth}
\end{table}

Despite the similarity in the arguments in these diverse competitive systems, the extensive evidence, and the profound implications for societal welfare, we lack a coherent theory of the underlying processes. We believe this has two main reasons, one empirical, one theoretical. A general empirical problem is posed by an inherent complexity/ambiguity gradient between proxy and goal. Specifically, the aspects of the societal goal which are most easily quantified are most likely to be captured in the proxy, biasing corruption toward whatever is difficult to quantify. Accordingly, much of the evidence suggesting a corruption of practices tends to be qualitative, voiced in interviews, editorials and surveys \cite{Fochler2016, Baker2016, Hicks2015, Wilsdon2015}. Additional goal aspects may be costly to assess or only become apparent in the long term. Such aspects can be quantified as alternative proxies and will remain unaffected by corruptive pressures as long as they play a minor role in competition. For instance scientific reproducibility, long term costs of excessive financial risk taking and environmental degradation play a minor role in competition and thus provide a quantitative measure of corruption \cite{OpenScienceCollaboration2015, Benabou2016}. Additional potential quantitative measures are specific readouts of corruption, such as retraction counts \cite{Brembs2013} or small sample sizes \cite{Smaldino2016}. The current revolution in data-science promises to provide ample testing ground for the present theory \cite{Fire2018}. However, due to the complexity of societal systems any individual study providing evidence in any of these domains remains difficult to interpret in isolation. 

The second, theoretical, challenge to a cohesive theory of proxy-based competition is its necessarily trans-disciplinary nature. We identify three fundamental questions such a theory must address. These are traditionally treated in separate disciplines. Importantly, all three questions arise directly from the core problem of proxy-based competition in societal systems (Table \ref{tab:tab1}):
\begin{enumerate}
\item What is the informational relation between proxy measures and the societal goal in the space of all possible practices/actions? (complex systems and control theory; information economics)
\item How do individual agents make decisions, given potentially conflicting information and value dimensions? (multitasking theory; contest theory; behavioral/neuro- economics; behavioral ethics)
\item How do inter-individual mechanisms such as selection/ cultural evolution affect the system?  (sociology; cultural evolution theory)
\end{enumerate}

We separate these questions for narrative convenience and conceptual clarity, though they are complexly interrelated. While the first and third questions address primarily the system level the second question concerns psychological decision mechanisms. In the following, we describe a simple agent-based computational model, constructed to integrate and formalize confluent insights across these disciplines and feedback loops between levels. The overall goal is to determine the principal components necessary to capture the processes described above and explore possible formal specifications. In the process, we outline the tenets of a unified theory of what we will call \textit{Proxyeconomics}, which applies to any proxy-based competitive system maintained by society to serve an abstract goal. Briefly, our model results suggest that any such system approaches a relatively stable equilibrium level of corruption determined by the information captured in the proxy and the strength of a putative intrinsic incentive 'to do your job well'. Moreover, the dynamics of corruption, or corruptive pressure, is governed by the intensity of competition and potentially the complexity of practices. Additionally, at the meso-scale, we find a range of incompletely understood but consistently observed effort choice patterns, described in the experimental contest literature \cite{Dechenaux2014}. For instance, effort-incentivization appears to be hill shaped with respect to competition, i.e. effort is highest for intermediate, rather than very mild or very intense, competition. 
In the following, we will first sequentially introduce and motivate the model components. We will then present detailed results on model behavior and robustness. Next, we will discuss these results including the model's limits, wider implications and possible extension. Finally, we will briefly discuss a number of additional implications of the theory of proxy-based competition including several psychological, economic, political and moral aspects. 

\section{The model}
We developed the model in order to capture the following general notions, each representing convergent insights by the respective disciplines, in the most parsimonious  yet plausible way.
\begin{enumerate}
\item The proxy can be viewed as the `objective function' of an optimization problem in a complex adaptive system \cite{Manheim2018A,Amodei2016,Flake1998}. Therefore there will always be some informational difference between proxy measures and societal goal (likely proportional to the complexity of the societal system).
\item Individuals are motivated not only by egoistic interests but also by an intrinsic moral motivation. \cite{Hochman2016,Pittarello2015,Chugh2016,Holmstrom1989,Benabou2016}. Furthermore, in competition, a powerful desire to (professionally) survive suggests some form of (proxy-based) reference dependence, well captured by a sigmoidal utility function \cite{Kahneman1979,Mishra2014,Delgado2008,Nieken2008,Gonzales2017}.
\item Variability in culturally/professionally transmitted practices and selection through competition introduces cultural evolution at the system level \cite{Smaldino2016}.
\end{enumerate}

Importantly, there will be interactions including feedback loops within and between individual and system level. The information affecting individual decisions will depend on the information at the system level. The competition experienced by individuals relates to the competitive selection at the system level. Finally, and most importantly, proxy performances affect outcomes primarily as a relative measure, i.e. by their relation to other agents' proxy performances. This implies feedback loops for i) psychologically driven proxy performance, ii) selection driven proxy performance and iii) between the two processes. An agent-based modeling approach is ideally suited to capture these notions including the feedback loops within and between levels.
 
The model draws from information/control theory and economic principal-agent theory by viewing a societal system as an information processing device or principal, respectively (see detailed model description and discussion\cite{Manheim2018A, Holmstrom1991}). A neuroeconomically plausible (utility and reference based) effort incentivization mechanism \cite{Glimcher2016} is modeled using concepts from prospect theory and multitasking theory \cite{Kahneman1979,Holmstrom1991,Mishra2014, Benabou2016}. Simultaneously, the processes described by Campbell's Law are modeled as a form of cultural evolution, on a slower but overlapping timescale \cite{McElreath2007,Smaldino2016}. The combination of these approaches allows us to investigate the tension between positive and negative effects of competition. Specifically two positive effects, namely effort incentivization and signaling/screening of talent, cooccur with negative effects, namely incentives to game the system and selection of corrupt practices. The model furthermore provides a way to probe the power of `intrinsic motivations' to bound a long term evolution towards corrupt practices, as recently suggested by Smaldino and McElreath \cite{Smaldino2016}. 

Briefly, we model a societal system, in which utility maximizing agents repeatedly compete with each other based on individual proxy-performances. Individual performances result from both the individual's effort and the orientation of their practice toward the proxy. The system is characterized by the following: The amount of corruptible proxy information, parametrized as the goal angle ($ga$, competition ($c$), parametrized as the fraction of losers; cultural evolution, parametrized through a practice mutation amplitude ($m$) and selection pressure ($sp$). Intuitively, $ga$ and $m$ reflect the complexity of the societal system while $c$ and $sp$ respectively reflect the psychologically experienced and realized intensity of competition. Individual utility-maximizing agents are characterized by  talent ($t$) and a parameter determining the relative power of an intrinsic moral incentive toward the societal goal over the extrinsic competitive incentive ($gs$). Finally, the individual agent's properties of effort ($e$) and practice orientation ($\theta$) represent the main outcomes of interest, informing on i) the emergent corruption and ii) the systems overall efficacy. In the following we will sequentially introduce our model specifications and parameters in the three building blocks mentioned above, beginning with a short section on competition and model time course.

\section{Detailed model description}
\subsection{Competition and model time-course} 
Competition is a central variable, which is likely to impact individual decisions as well as system-level selection. For a given population, we define competition simply as the fraction of `losing' agents $c \in [0,1]$, i.e. agents in danger of (professionally) dying due to insufficient proxy-performance at a given time point. Professional death (briefly refereed to as death) refers to the permanent elimination from the respective system. For instance, in intense competition ($c=0.9$) an agent has to be in the top 10\% of proxy-performers to be a `winner'. This then affects both the psychological decision mechanism and system-level selection as described below.\\ 
The model proceeds in time steps, each of which is subdivided into a decision phase and an evolution phase. In the decision phase all agents are drawn in random order to adjust their efforts to maximize utility given their individual properties ($\theta, t$. see below) and the currently observed competitive rank (of proxy performance). In the evolution phase, losing proxy performers are subject to stochastic death, and are instantaneously replaced by offspring from winning proxy-performers, such that the population size stays constant. Together, this models a competitive system in which agents make their decisions based on the observed proxy-performances of all other agents, but are unsure about the exact time frame of competition, i.e. to which degree other agents might still change their proxy performance before a selection event might occur. The continuous repeated comparisons, occurring in consecutive time steps resemble what has been called `multi-contest tournament' \cite{Fu2015}. 

\subsection{The practice space - information content of the proxy}
\begin{figure}
\centering
\includegraphics[width=\textwidth]{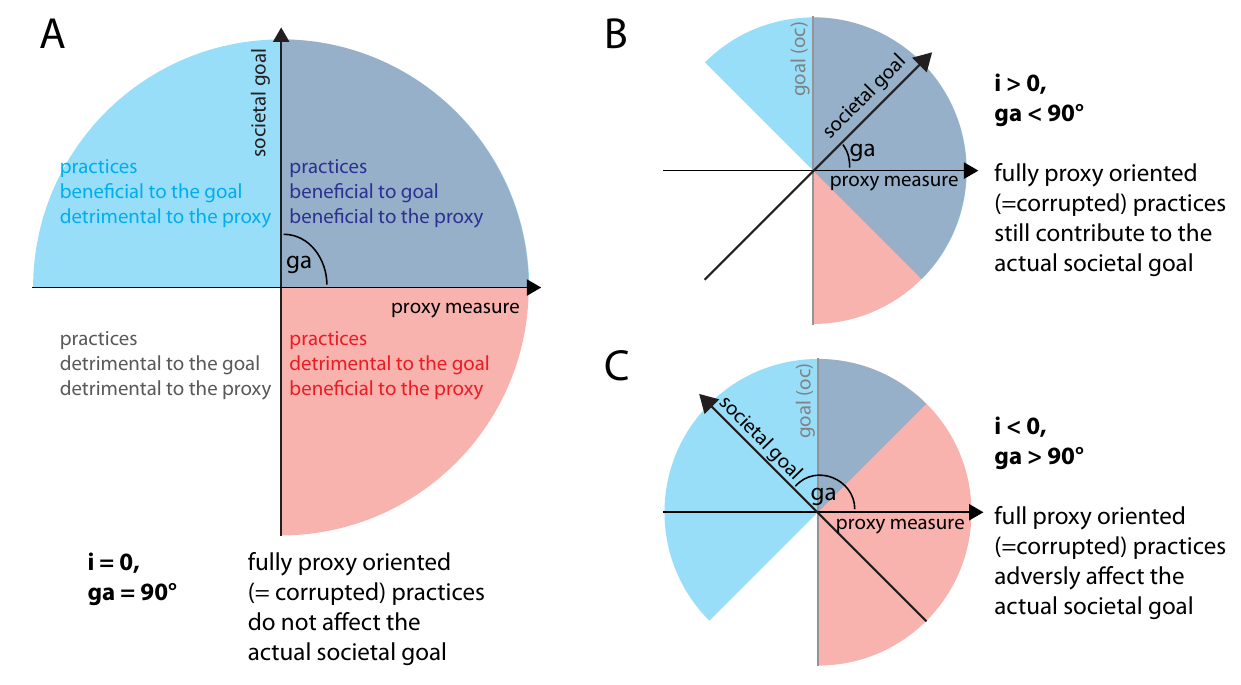}
\caption{\textbf{The practice space,} Individual practices may contribute to different degrees to the proxy measure and the societal goal, motivating the mapping of all practices to a two-dimensional practice space. Sections of the practice space which are beneficial/ detrimental to the proxy/goal are color coded (see colored labels in \textbf{A}). The degree to which the proxy captures true and incorruptible information about the societal goal can then be represented as the angle between the main axes, denoted goal angle ($ga$) with information $(i=cos(ga))$. Exclusively proxy oriented practices are located on the horizontal axis (proxy measure). Accordingly, when such (corrupted) practices are neither beneficial nor detrimental to the societal goal, then $ga = 90^\circ$ and $i = 0$ \textbf{(A)}. Presumably, in most cases the proxy captures sufficient incorruptible information about the goal such that even even fully proxy oriented practices contribute to the societal goal ($0^\circ < ga < 90^\circ; 1 > i > 0$  \textbf{(B)}. However, full proxy orientation may also lead to negative outcomes for the actual societal goal ($90^\circ < ga < 180^\circ; 0 > i > -1$  \textbf{(C)}. When $ga \neq 90^\circ$ the goal component orthogonal to the proxy is labeled `goal (oc)'.}
\centering
\label{fig:Fig1} 
\end{figure}

We conceptualize the informational relation between proxy and goal with respect to the value-creation associated with specific cultural practices. A cultural practice is defined as a specific pattern of actions, that can be associated with potentially differing relative contributions to the societal goal and the proxy (similar to e.g. \cite{Smaldino2016, Pluchino2011, Benabou2016} in the fields of science, politics and economics, respectively).
In the main model the cultural practice is predominantly learned and transmitted within a given cultural entity (laboratory, company, ...), but we later also explore outcomes if practices are subject to individual agency. Four fundamental groups of practices can be intuitively differentiated based on the degree to which they are beneficial or detrimental to the proxy and goal value (Fig~\ref{fig:Fig1}A). For instance, in any specific societal context, we may consider if practices exist that exclusively contribute to either the proxy measure or the societal goal. The existence of such practices motivates the dimensional reduction of the practice space to two dimensions representing the proxy and the goal (as in economic multitasking \cite{Holmstrom1991, Benabou2016}). The degree to which the proxy captures true and incorruptible information ($i$) about the societal goal can then be represented as the angle between the main axes (goal angle; $ga \in [0^\circ$, $180^\circ]$), where information $i \in [-1,1]$ is given by Eq. \ref{eq:Information}. 

\begin{equation}
\label{eq:Information}
i = cos(ga)
\end{equation} 

Intuitively, information measures the effect on the societal goal when practices are exclusively oriented towards the proxy. A good proxy ($0<i<1$) captures sufficient information about the societal goal, such that even fully corrupted practices produce positive outcomes for the actual societal goal (Fig~\ref{fig:Fig1}B). In contrast, when full proxy orientation produces negative externalities which outweigh the beneficial effects for the societal goal, this is captured by $-1<i<0$ and $90^\circ < ga < 180^\circ$ (Fig~ \ref{fig:Fig1}C). Externalities beyond the dimensions described above, i.e. that are independent of both proxy and goal (e.g. leisure time), are not considered in the current model. We reason that $i$ and $ga$ relate directly to the complexity of the system with $i$ tending toward $0$ and $ga$ toward $90^\circ$ as complexity and, with it, the number of failure modes increases \cite{Manheim2018B}.
 
\subsubsection{The proxy}
We can most easily think of the proxy as a simple scalar metric, such as the journal impact factor or quarterly profits. More generally, the proxy should be thought of as an arbitrarily complex regulatory model \cite{Conant1970}, embodied in the societal mechanisms designed to collect information and transform it into competitive rankings. Arbitrarily complex, here, explicitly includes mechanisms relying on expert judgement \cite{Strathern1997}, as long as the outcome is a competitive ranking. In analogy to the economic revealed preference approach, we can think of the proxy as the set of attributes that factually determine selection within the competitive system (the revealed preference of the system/mechanism). In analogy to the machine learning and artificial intelligence literature, it is helpful to think of the proxy as an objective function \cite{Manheim2018A}. In all these analogies, the societal system is conceptualized as an information processing device, collecting complex input information and converting it to an actionable output metric (competitive rankings). The boundary case of a perfect regulatory model maps to a goal angle of $ga = 0^\circ$.

\subsubsection{The goal}
The societal goal is the arbitrary consensus set of properties society wants to achieve/regulate (see examples in Table \ref{tab:tab1}). Though this is difficult to precisely define in any specific context, it is important to bear in mind that such a consensus set is implied by the fact that society maintains an artificial competitive system in the first place. Practical and information-theoretic considerations further help to predict the ways a goal is likely to differ from the proxy \cite{Conant1970, Manheim2018A, Amodei2016}. This can allow a relational or negative definition (e.g. a preponderance of irreproducible publications may not further the societal goal). Note that, in the present model, we define the societal goal as only those goal-aspects, which are privately known to individual agents. This definition of the societal goal may be most productive, since aspects of the goal unknown at the individual and the institutional level may be particularly difficult to address. In the present model, individuals are assumed to have superior knowledge as to the precise profiles of their actions and their implications. Indeed, there are numerous reports of individuals in competitive systems who state that competition is impeding their ability to act according to the societal goal. This directly implies private knowledge of implications for the goal which are not captured by the proxy (for instance see \cite{Baker1992, Benabou2016} for examples from science and banking, respectively). Similarly, a substantial portion of the general population personally feels that their whole job (which is the result of competitive market mechanisms) does not contribute to societal welfare in any way \cite{Graeber2018}.

\subsection{Individual choice - multitasking}
\begin{figure}
\includegraphics[width=\textwidth]{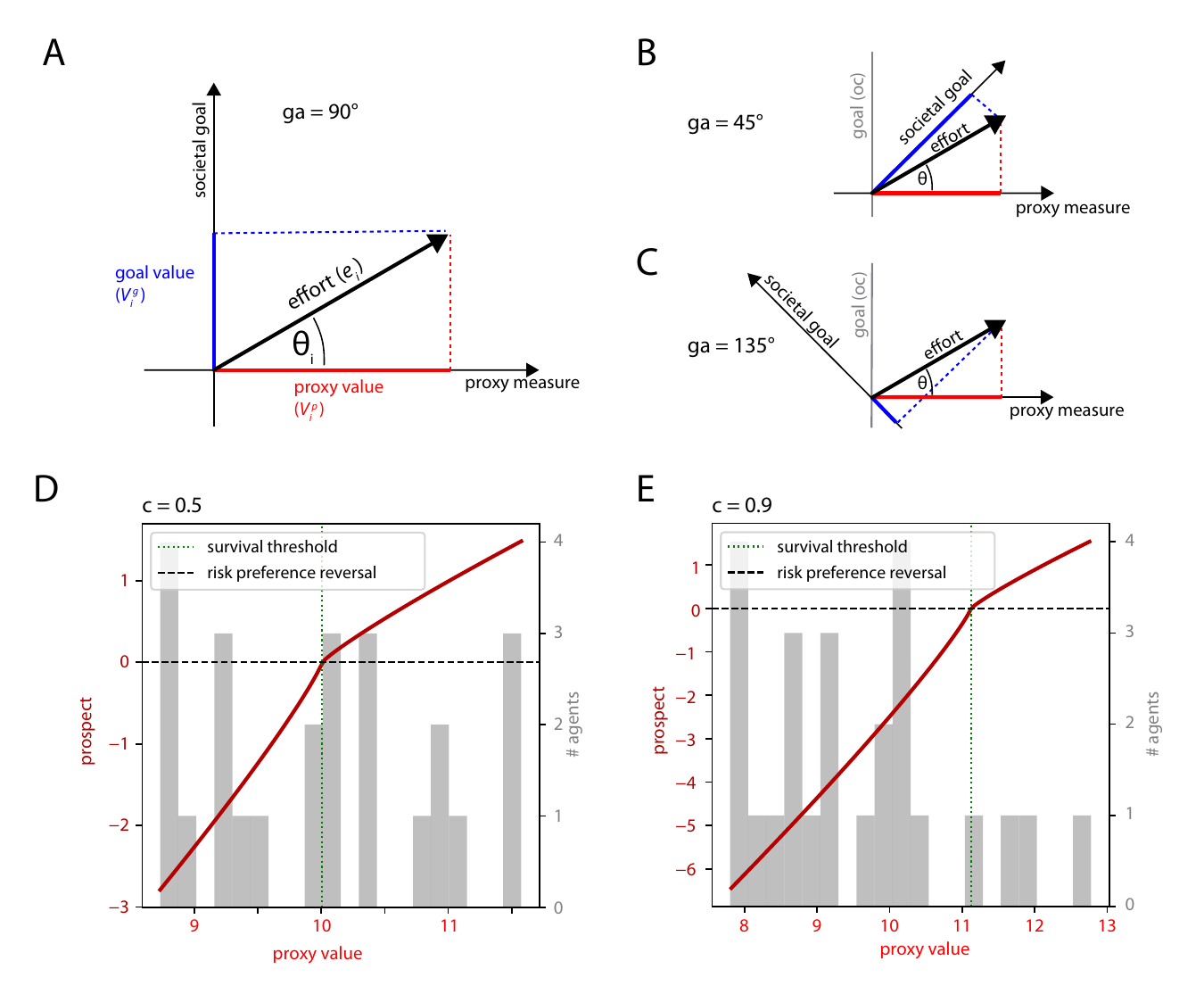}
\caption{\textbf{Agent decision model,} Agents derive utility both from contributing to the societal goal and performing well according to the proxy measure (blue, red respectively). An individual agent (index i) produces goal value ($V_i^g$) and proxy value ($V_i^p$) as determined by the practice angle ($\theta_i$) and effort ($e_i$), simply as the projection of the effort vector onto the main axes (eqs. \ref{eq:GoalValue} and \ref{eq:ProxyValue}; \textbf{A-C}), where the goal angle ($ga$) is the angle between the main axes (see Fig~\ref{fig:Fig2}). The utility derived in the proxy dimension (denoted prospect) depends not on the absolute proxy value but on the relative ranking of proxy values and the survival threshold according to eqs. \ref{eq:ProxyUtility} and \ref{eq:LossAversion}. Panels \textbf{D, E} show illustrative prospect functions (dark red) for competition $c=0.5$ and $c=0.9$ respectively. The survival threshold (vertical spotted line in D, E) is the salient reference proxy value separating losers from winners, where competition ($c$) denotes the fraction of losers. Accordingly, for $c=0.5$ or $c=0.9$, the survival threshold is the $50^{th}$ or the the $90^{th}$ percentile of the population distribution of proxy values respectively (grey histograms in D, E respectively).}
\label{fig:Fig2} 
\centering
\end{figure}

The system-level definition of the practice space now allows us to intuitively capture the notion of proxy-orientation of a given agent's practices by their `practice angle' ($\theta$; Fig~\ref{fig:Fig2}A-C). The agents effort level is assumed to be the length of a vector along this practice angle and the true contributions to proxy and goal value are the projections onto the main axes. This captures a positive effort to output relation, where only the type of output depends on practice orientation. To allow both, an intrinsic (goal) and a competitive (proxy) incentive to elicit effort, we adapt an economic principal-agent multitasking model (Fig~\ref{fig:Fig2}A). Note that in contrast to traditional multitasking models, we here allow practice angles to be established either through individual choice, or through a process of cultural evolution (i.e at the system level). The practice angle of an agent reflects the average outcome of all the different actions and choices she makes in an individual time step, i.e. it may include strategies such as cheating or sabotaging a fraction of the time, which would translate to a lower average $\theta$.
 
Accordingly, the effort of the $i^{th}$ agent $e_i \in \mathbb{R}_{\geq0}$ is modeled as the length of the vector with orientation $\theta_i$ and the resulting proxy and goal values are the projections onto the main axes capturing the practice dependent trade-off (Eq \ref{eq:GoalValue},\ref{eq:ProxyValue}). 
 
\begin{equation}
\label{eq:GoalValue}
V^g_i(\theta_i,e_i) = cos(ga-\theta_i) \cdot e_i 
\end{equation}

\begin{equation}
\label{eq:ProxyValue}
V^p_i(\theta_i,e_i) = cos(\theta_i) \cdot e_i 
\end{equation}

The utility derived from the goal performance is simply the goal value multiplied by a constant ($gs$), determining the relative psychological strength of an intrinsic, moral incentive toward the societal goal (Eq \ref{eq:GoalUtility}). Given that goal in our settings refers to aspects known to agents, $gs$ can be seen as a product of i) goal valuation and ii) individually available goal information.

\begin{equation}
\label{eq:GoalUtility}
U^g_i(V^g_i)= gs \cdot V^g_i
\end{equation}

\subsection{Individual choice - competition as incentive}
The utility derived from the proxy value, denoted `prospect' ($U^p_i$), is determined through competition (Fig~\ref{fig:Fig2}D, E). It depends on the relation of the own proxy value $V^p_i$ and the proxy value required for professional survival (the survival threshold, $ST$). The survival threshold is the proxy value which separates winners and losers for a given level of competition. For instance $c = 0.9$ indicates that the survival threshold is the proxy value at the $90^{th}$ percentile of the distribution of all proxy values, i.e. an agent has to be in the top 10\% proxy performers to be a winner. The survival threshold is assumed to be the salient reference point with respect to which agents evaluate the utility of their own proxy value. Accordingly, they will derive negative/positive utility if they are below/above the survival threshold.  
In the first version of this manuscript/model we have considered a prospect function based on a Gaussian uncertainty distribution around the survival threshold (based on \cite{Nieken2008, McDermott2008, Mishra2014}). Though the main results were similar (see manuscript history on arXiv), this function was not scale invariant with respect to proxy value. Therefore we here choose a similarly sigmoid, but scale invariant prospect function, namely that of cumulative prospect theory (Eq \ref{eq:ProxyUtility} \cite{Tversky1992}). 

\begin{equation}
\label{eq:ProxyUtility}
U^p_i(V^p_i,ST,\epsilon)= (V^p_i- ST)^{0.88}
\end{equation}

Additionally, we assume agents are loss averse, i.e negative prospects are multiplied by 2.25 (Eq \ref{eq:LossAversion}, \cite{Tversky1992}).

\begin{equation}
\label{eq:LossAversion}
U^p_i \mapsto 
\begin{cases}
	U^p_i ,&\text{if } U^p_i\geq 0\\
	U^p_i \cdot 2.25,&\text{otherwise}
\end{cases}
\end{equation}

In a subset of models, we additionally probed a simple step function as prospect function [-1,1]. 
Each agent $A_i$, when she is drawn, chooses her effort level to maximize her individual utility $U_i$ given the observed current proxy performances of all other agents (Eq \ref{eq:AgentUtility}),
  
\begin{equation}
\label{eq:AgentUtility}
U_i(U^g_i,U^p_i,cost) = U^g_i + U^p_i - cost(e_i,t_i)
\end{equation}

where effort cost is given by:

\begin{equation}
\label{eq:EffortCost}
cost(e_i,t_i)= e_i^2/t_i
\end{equation}

Here $t_i$ is the talent of the agent, i.e. a constant determining the relative cost of effort independent of $\theta$. Agents are given variable talent according to $\mathcal{N}(10,tsd)$ such that the effort to proxy-performance relation depends on both individual practice orientation and talent. This allows us to capture beneficial effects of the proxy as signaling/screening device for talent, as well as detrimental effects (Campbell's Law). Note, that the the latter could similarly be described as signaling/screening of corrupt practices, and our model allows beneficial and detrimental mechanisms to act simultaneously.
In order for individual agents to compute the complex maximization problem described above, we assume they consider a limited range of effort adjustments, namely [-10, -5, -1, -0.5, -0.1, 0, 0.1, 0.5, 1, 5, 10]. The rationale behind this is that the agent intuitively judges what might happen if she changes her effort just a little or a lot but lacks the computational capacity for perfect maximization. Nevertheless, if the system is stable enough, she will iteratively approach the optimal effort. Note, that alternative effort choice lists (e.g. -10 to 10 in 0.1 increments) did not change model outcomes but dramatically increased the computational burden of agents (and model). Furthermore, the specific range (-10 to 10) simply covers the magnitude of plausible effort changes in the system arising from the arbitrary mean talent choice of 10, and choosing a larger range did not change outcomes. In a subset of models, practice agency was introduced by letting agents maximize utility over the effort test-list for each of a range of practice-adjustments in an angle-list of [-5°, -1°, 0°, 1°, 5°], following the same rationale.

\subsection{Selection and evolution}
When all agents have been randomly drawn to adjust their efforts (and potentially practice), the ranking of proxy-performances is reassessed. Each potential loser ($V^p_i < ST$) is then subject to professional death with probability = selection pressure ($sp  \in [0,1]$). Accordingly, $sp$ determines the approximate number of steps one can afford to be a loser before being actually removed, relating \textit{experienced} to \textit{realized} competition. Upon each death a position opens up, which allows a randomly chosen `winning' proxy-performer ($V^p_i > ST$) to professionally reproduce, passing her practice angle on to her offspring. \\
During practice inheritance, $\theta$ mutates stochastically, such that $\theta_{\textit{offspring}} = \theta_{\textit{parent}} + \mathcal{N}(0,m)$. We reason that the magnitude of the mutation rate $m$ is proportional to the complexity of the societal system. Large mutation rates reflect potentially large and frequent effects of minor practice changes on proxy and goal values driven by i) an increasing number of nonlinear interactions in more complex systems and ii) the combinatorial explosion of possible action to outcome mappings with increasing complexity \cite{Manheim2018B}. In this context it is important to remember that the practice space represents a dimensional reduction to the orthogonal components of proxy and goal. Accordingly, mutations can be thought of as comprising arbitrary changes in additional independent dimensions, for instance animal welfare in meat production (assuming animal welfare is not considered a societal goal of the industry).  

\subsection{Implementation/ model development/ parameter choice}
The model was implemented in the agent-based modeling framework `Mesa0.8' in Python3.6 and run on a standard Windows7, 64bit  operating system. The full code to run the model and generate figures is attached as supplemental material.\\
Table \ref{tab:Parameter space} shows an overview of the explored parameter space with the parameters shown here highlighted in boldface (also see Fig~\ref{fig:Sensitivity}). A number of additional parameters were varied to probe robustness, but lead to no change, and will be reported at appropriate locations throughout the manuscript. The model was initialized with population size N (typically 100), where every agent received a practice angle $\theta$ drawn randomly from a uniform distribution between proxy and goal axes $\mathcal{U}(0,ga)$ and initial effort 0. Model runs were typically repeated 10 times for each level of competition to obtain measures of the mean and spread of system behavior. In each model run, all agents in a population compete against each other via their proxy performances as described above.

\begin{table}[!ht]
\begin{adjustwidth}{-2.25in}{0in} 
\centering
\caption{\textbf{Parameter space}}
\begin{tabular}{|>{\bfseries}p{2cm}||p{2cm}|p{5cm}|p{6cm}|}
\hline
Parameter & \textbf{Base Value(s), Range} & \textbf{Description} & \textbf{Main effect} (if parameter is increased)\\ 
\thickhline
\multicolumn{4}{|c|}{equilibrium determining parameters}  \\  
\hline
goal angle ($ga$) & \textbf{45}, \textit{90, 135}, 0 -180 & angle ($^\circ$) defining the amount of corruptible proxy information & hill shaped effect on equilibrium corruption, the optimal level of competition decreases \\ \hline
goal scale ($gs$)&  \textbf{1}, 2, 0-10 & scaling factor of psychological goal valuation (relative to experienced prospect value) & increased effort, decreased equilibrium corruption\\	\thickhline
\multicolumn{4}{|c|}{dynamics determining parameters}  \\  \cline{1-4}	
competition ($c$)&  \textit{0.3, 0.6}, \textbf{0.9}, 0.1-0.9 & competitive pressure, i.e. the fraction of potential losers per round & complex effects on effort, evolution and utility (see main text), increased speed of convergence to equilibria\\ \hline
selection pressure ($sp$)&  \textit{0.001}, \textbf{0.1}, 0.001-1  & probability of death for each losing agent in each round & increased speed of convergence to equilibria\\ \thickhline
\multicolumn{4}{|c|}{parameters affecting variability}  \\  \cline{1-4}	
\hline
talent standard deviation ($t_{sd}$)& \textbf{1}, 0-6 & standard deviation of talent within the population & increased effort spread \\ \hline
population size ($N$)& \textbf{100}, 10-500 &  number of agents in the system & decreased variability over models \\ 	\hline
practice mutation rate ($m$)& \textbf{2}, 0-30 & standard deviation ($^\circ$) of practice angle mutations during inheritance & increased practice variability and, if the equilibrium practice is outside the initialization range, speed of convergence to equilibria \\ \hline
\hline
\end{tabular}
\begin{flushleft} Overview of the parameter space. During sensitivity analysis individual parameters were varied in the specified ranges holding all other parameters at the base values. Presented base values are in boldface (Fig~\ref{fig:Sensitivity}, \ref{fig:Robustness}). The remaining base values were additionally probed.
\end{flushleft}
\label{tab:Parameter space}
\end{adjustwidth}
\end{table}

\section{Results}
We will first introduce the model using an exemplary parameter set and short timescale, emphasizing the emergent patterns of effort and practices, in order to demonstrate the surprising scope of overlap with empirical observation from the experimental contest literature. We will then demonstrate the main result of competition-induced equilibrium corruption at a longer timescale. Finally we will present the results of a detailed sensitivity analyses demonstrating the robustness of equilibrium corruption and revealing the parameters governing it.\\
To introduce the model we consider a system in which the proxy contains some incorruptible and some corruptible information about the societal goal ($ga = 45^\circ; i = 0.5$). Goal scale ($gs$) is set to one implying comparable weighting of proxy and goal incentive. Selection and mutation are mild ($sp=0.1, m=2^\circ$), implying a $10\%$ probability of death for losers at each time step and small changes of practice orientation during inheritance. At initialization each agent receives a random practice angle between full corruption and full goal orientation $\theta =\mathcal{U}(0,ga)$ and normally distributed talent $t = \mathcal{N}(10,1)$.

\subsection{Iterative effort choice - competitive incentivization and discouragement}
We begin by describing the emergent behavior resulting from iterative effort choice. In brief, at each time step, agents in random order adjust their effort given the noisy observations of proxy performances of competing agents and their personal parameters (prospect function, practice angle and talent). This simple contest specification produced a range of behaviors observed in experimental economics including i) optimal effort incentivization at intermediate levels of competition, ii) a discouragement effect and iii) effort bifurcation \cite{Dechenaux2014}. Importantly, we did not consider or anticipate any of these effects during model design, but rather strove to create the simplest effort choice algorithm applicable in a step wise agent-based model with plausible preferences, information acquisition and computation of individual agents.
The algorithm produced a range of interesting individual effort trajectories, highly dependent on competition (Fig~\ref{fig:Fig3}A, four exemplary model runs ranging from very low to very intense competition). Note that individual agent properties (practice orientation $\theta$ and talent $t$) are identically distributed between competition levels. However, in higher competition, individual agents are forced to increase effort to cross a higher survival threshold (the competitive cutoff between winners and losers). This in turn affects the population distribution of proxy values leading to a positive feedback loop and higher effort levels in higher competition. When marginal effort cost begins to outweigh the expected utility gain, individual agents may stop increasing or begin to decrease effort. Indeed, due to the flattening of the prospect function when the observed survival threshold (given other agents performances) is far beyond the own proxy performance, agents may become `discouraged', i.e. decrease effort (Fig~\ref{fig:Fig3}A, e.g. black arrows). In such cases the prospect of winning becomes unrealistic with acceptable effort expenditure, such that agents opt to save effort cost. Discouragement becomes more frequent in more competitive systems, leading to an eventual reversal of the competition to effort relation for some agents, as observed empirically \cite{Harbring2008, Orrison2004, Bradler2016}. Increasing prevalence of discouragement with higher competition also entails an increase in variance of effort at the population level. At the individual level, agents similarly show more variable effort over time in higher competition. This competition to effort-variance relation is highly robust, notably even without agent heterogeneity (not shown), a testable prediction contrasting our model with previous contest models \cite{Dechenaux2014}. Notice, how some agents progressively increase effort in order to compete, but eventually become discouraged (Fig~\ref{fig:Fig3}A, black arrows). The resulting decrease in the survival threshold may in turn provide other agents with a prospect of winning, such that they can gain utility by increasing effort (Fig. \ref{fig:Fig3}A, grey arrow).
To the best of our knowledge, this is the first neuroeconomically plausible iterative effort choice algorithm which provides a mechanistic account of how equilibrium effort is approached in contest settings (see also \cite{vandenBos2013}). Notably, it reproduces a remarkable range of empirically observed but incompletely understood phenomena concerning both static effort distributions and effort trajectories and is amenable to experimental verification.\\
Additional to these psychologically driven effects, competition determines selection by defining the proportion of losers. Agents with insufficient proxy-performance (losers) are subject to stochastic professional death with probability $= sp$, where death means permanent elimination from the competitive system (indicated by white rectangles; Fig~\ref{fig:Fig3}A, e.g. white arrow). Upon a death, the free slot is immediately filled up by the offspring of a random winning agent. Given, that competition determines the proportion of losers and losers have uniform probability of death, selection events increase proportionally with competition. Note that this specification keeps the population size constant, modeling a societal system with fixed size and resource consumption where increased competition is realized through an increased throughput of new agents.

\begin{figure}
\centering
\includegraphics[width=\textwidth]{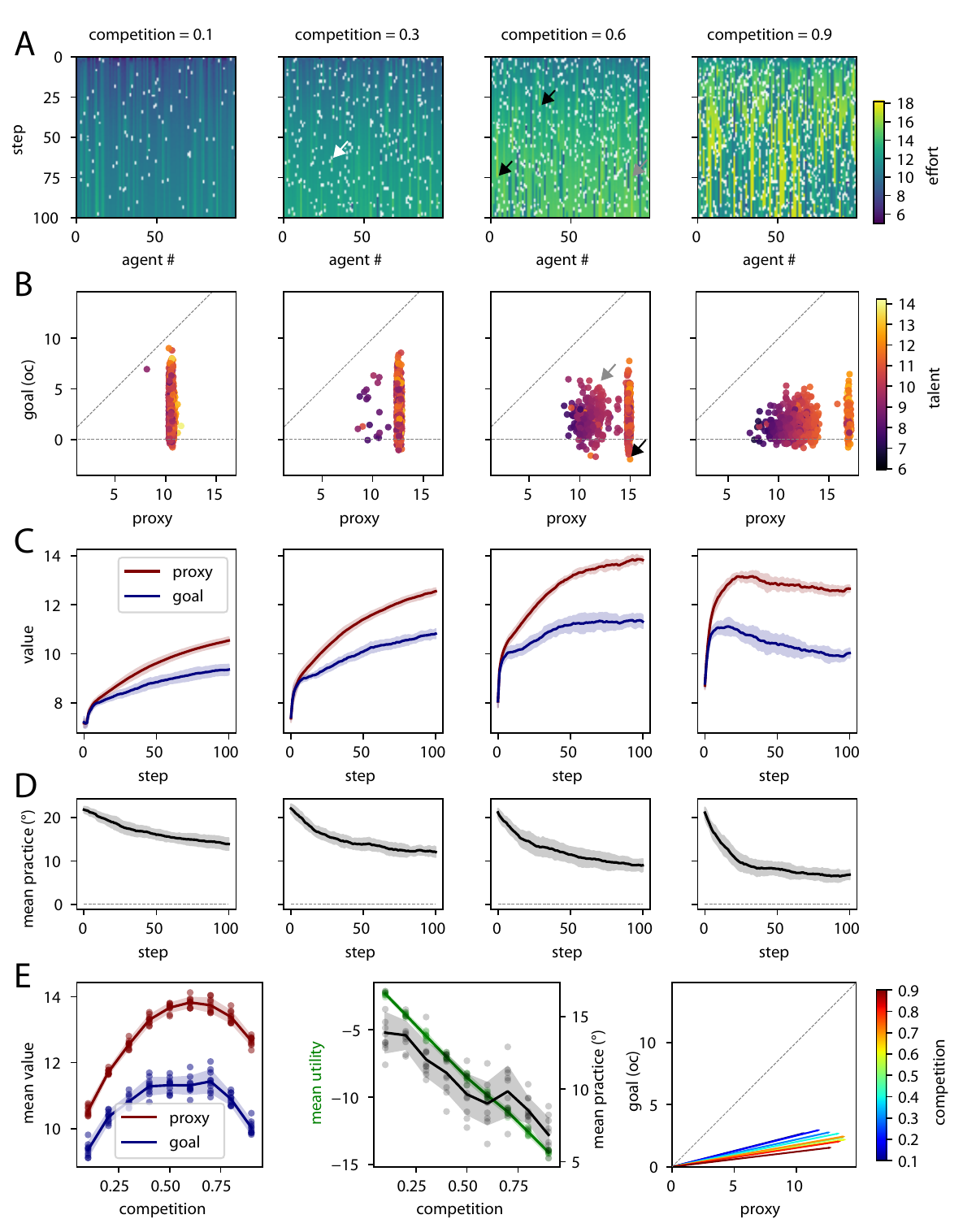}
\caption{\textbf{Short term dynamics}}
\label{fig:Fig3}
\centering
\end{figure}  

\begin{figure}
\centering
\contcaption{\textbf{Short term dynamics} (100 time steps), with $ N=100, ga=45^\circ, \theta_{initialization}=\mathcal{U}(0^\circ,ga), gs=1, t= \mathcal{N}(10,1), sp=0.1, m=2^\circ$. Data are collected from 10 model runs per competition level ($c$). A-D show four levels of competition (major columns) while D shows $c=[0.1,0.2,...,0.9]$. A and B show individual agents while C-E show population means from repeated model runs as data points. \textbf{A) Agent Dynamics,} Each subpanel shows effort over time for each of the 100 agents (random run for given $c$). Columns within subpanels represent individual agents or positions. White squares indicate death/birth events (two example highlighted with white arrows). Black arrows indicate examples of agents becoming discouraged by competition. Grey arrows indicate examples of agents increasing effort to gain the prospect of winning.  \textbf{B) Agent Values,} Realized proxy and goal values of each agent at the 50th time step. Values are the result of the current chosen effort given the individual agents practice angle (compare Fig \ref{fig:Fig3}B). Each data point represents an agent (agents accumulated from 10 runs per panel). Talent is color coded. The black and grey arrows indicate highly proxy or goal oriented agents, respectively. Note that the agents indicated by the grey arrow are outperforming those indicated by the black arrow concerning their goal performance, but not proxy performance. \textbf{C, D) Model Dynamics,} Model level dynamics of proxy and goal value (C) and the mean practice angle (D).  \textbf{E) Model Values,} Final (step 100) mean proxy and goal values (left) and utility and practice angle (middle) as a function of competition. Data on mean proxy and goal values can be projected back into the practice space (right) as in Fig~\ref{fig:Fig2}A. The shaded area around lines represents the standard deviation over model runs.} 
\end{figure}  

\subsection{Realized value in the practice space - simultaneous beneficial and detrimental signaling} 
The relative ability of individual agents to compete depends on two parameters, namely talent and practice orientation, both of which are not directly observable. Talent allows greater effort, independent of practice orientation. Accordingly, if practices are fixed, the proxy fulfills its intended role as signal of goal performance, allowing efficient screening by the competition. At the same time, however, high proxy performance may result from higher proxy orientation of the practice, implying wasteful or even detrimental signaling (Campbell's Law). In our model, both processes act simultaneously and can be assessed by visualizing realized practice-effort pairs of individual agents back into the practice space with color-coded talent (Fig~\ref{fig:Fig3}B). Here each data point represents the endpoint of the effort vector (as in Fig~\ref{fig:Fig3}A-D) of an individual agent at time step 100. The resulting goal and proxy performances correspond to the projections onto the main axes, as shown in Fig~\ref{fig:Fig3}B). Across all levels of competition, we observe a dominant effect of the competitive incentive on realized effort levels, as outcomes tend to organize in vertical lines, i.e. they cluster around a specific proxy value. Further analysis showed that the vertical line corresponded to proxy-values just above the emergent survival threshold for a given level of competition. Agents just below this threshold had either increased effort attempting to enter the win-domain or decreased effort further, to save effort cost. Accordingly, the sigmoidal prospect function, including loss aversion, drives effort bifurcation, particularly when judging effort by proxy performance.\\ 
Also across all levels of competition, but most prominently for high competition, we observe a beneficial signaling/screening effect of the proxy as higher talent translates to higher proxy performance. We furthermore
Another interesting emergent phenomenon is a competition dependent reversal of relative effort expenditure by agents with high practice angles. In low competition (Fig~\ref{fig:Fig3}B, $c = 0.1$, leftmost panel) agents with high practice angles are forced to put in extra effort, but are still able to compete, leading to particularly high relative contributions to the societal goal. By contrast, when competition is intense (Fig~\ref{fig:Fig3}B, $c = 0.9$, rightmost panel) agents with more goal oriented practices can no longer compete on the proxy scale and are preferentially discouraged, even if they have high talent. Notice, that the emergent practice-effort realizations cover several qualitatively distinct domains across the practice space, where observable proxy performance only partially predicts unobservable goal performance. Some agents are peak proxy performers, while only moderately contributing to the actual societal goal (Fig~\ref{fig:Fig3}B, black arrow). Simultaneously, some highly talented agents are `losing' in proxy-competition, while actually outperforming many `winners' regarding the actual societal goal (Fig~\ref{fig:Fig3}B, grey arrow). More generally, the model predicts that for any observed proxy performance and at any time point, there is a mix of agents with highly varying degrees of proxy orientation. Thus our model intuitively captures the simultaneous effects of beneficial and detrimental signaling, which we believe to invariably arise in any proxy-based competition.

\subsection{Short term system dynamics}
Preferential discouragement of goal oriented agents might translate to a divergence between proxy and goal performance at the system level. To visualize variability at the system level, the following data are displayed as mean and standard deviation over model runs where each run is represented by the population mean of its agents. Indeed, mean proxy and goal performance of populations of agents show distinct competition dependent dynamics (Fig~\ref{fig:Fig3}C). Note an initial phase (within the first 10 steps), visible as a steep rise in value as effort first adjusts to an initialization independent level (models initialized with effort levels of 0, 1, 10 or 20 all converged to the same values during this initial phase). Subsequently, value may progressively increase (Fig~\ref{fig:Fig3}C, $c=0.1$ to $0.6$) or decrease (Fig~\ref{fig:Fig3}C, $c=0.9$) due to feedback loops between individual effort and survival threshold. More intense competition also leads to an increasing divergence between proxy and goal value (Fig~\ref{fig:Fig3}C). One driver of this effect is the preferential discouragement of agents with high goal orientation, implying an overall effort redistribution toward the proxy (a psychological mechanism of Campbell's Law). A second potential driver is a selective removal of goal oriented agents due to inferior proxy performance (a cultural evolution mechanism of Campbell's Law). The degree to which this happens can be assessed by viewing the dynamics of the population mean practice angle, which is independent of effort (Fig~\ref{fig:Fig3}D). With a goal angle of $45^\circ$, the initial uniform distribution of practices between proxy and goal leads to an initial population mean practice of $\theta =ga/2= 22.5^\circ$. However, as selection removes agents with high practice angles, replacing them with offspring from agents with lower angles, the population mean evolves toward the proxy. As expected, the speed of evolution towards the proxy is proportional to the intensity of competition since the quantity of selection events drives population change of practices. Thus, our model reproduces a central finding by Smaldino and McElreath \cite{Smaldino2016}, namely the the powerful corruptive force of proxy guided cultural evolution.

\subsection{System outcomes over competition}
To obtain a more complete view of proxy and goal performances we next plot the mean final values of this short period (step 100) as a function of competition (Fig.  \ref{fig:Fig3}E, left; data points represent model runs). Note that for both proxy and goal value, there is an optimal level of competition. No increases in goal performance are achieved by competition greater than $c\approx0.4$ while proxy-performance suggests $c\approx0.6$ is optimal.Very low competition implies low effort incentivization, leading to low value creation in both proxy and goal domains. As competition increases there is increasing effort incentivization but simultaneously proxy and goal values begin to diverge, due to the dual process of selective discouragement and evolution described above. Furthermore, at very high competition, an increasing fraction of discouraged agents, leads to decreasing mean effort.   
Another negative effect of competition is lower mean utility of the participating agents (Fig~\ref{fig:Fig3}E, middle panel). Though each agent individually maximizes utility, the necessity of being relatively better than other agents pushes the system to relatively high effort levels (and effort cost), as agents continuously attempt to break out of the loss frame \cite{Krakel2008}. Since the proportion of agents in the loss frame however ultimately remains fixed by competition this leads to decreasing utility. Notice, that we have not modeled positive utility from a flat wage component, since it does not affect the utility maxima of individual agents and thus system behavior. Adding a flat-wage utility (10 or 20 utils), increased utility by a constant independent of competition, linearly shifting the utility distribution upward (not shown), but not otherwise affecting model outcomes. Also notice however, that we feel it is not necessary to assume a participation constraint, as our agents model trained professionals and may accept substantial disutility rather than switching to a different profession, which may involve additional costly training. Instead, agents may become discouraged and cease participating due to a selection event (an endogenous participation constraint). Accordingly, low mean utility may be interpreted as either necessitating a higher flat-wage compensation or entailing a distressed agent population. Finally, we directly visualize the population mean practice-effort combinations within the practice space (Fig~\ref{fig:Fig3}E, right panel). Competition is color coded, such that the effect on both mean effort (vector length) and practice orientation (vector angle) are directly visualized (compare Fig~\ref{fig:Fig2}B and \ref{fig:Fig3}B). Thus, the model at the system level similarly captures the trade-off between positive and negative effects of competition, and reproduces the notion of an optimal level of competition resulting from both effort and practice dynamics \cite{Benabou2016, Manheim2018A}.

\subsection{Long term dynamics}
\begin{figure}
\centering
\includegraphics[width=\textwidth]{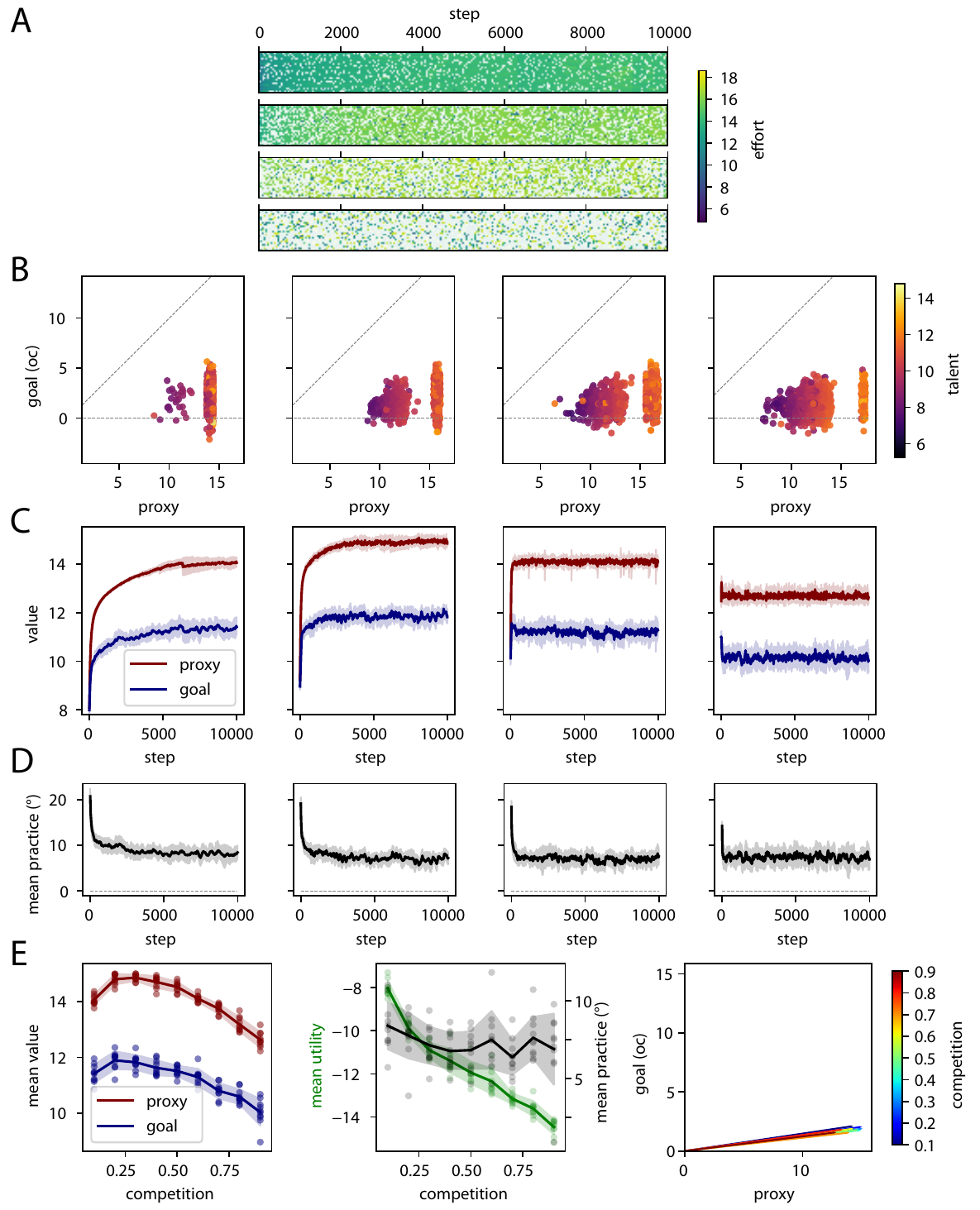}
\caption{\textbf{Long term dynamics,} (10000 time steps), with $N=100, ga=45^\circ, \theta_{initialization}=\mathcal{U}(0^\circ,ga), gs=1, t= \mathcal{N}(10,1), sp=0.1, m=2^\circ$. For panel descriptions please refer to Fig~\ref{fig:Fig3}. In (A) only a subset of agents and every $10^{th}$ time step is depicted.} 
\label{fig:Fig4}
\end{figure}

So far, we have explored the model dynamics in the short term (100 time steps). We reason this timescale is most relevant when assessing the dynamic effects of parameters, such as the competitive pressure. Notably, additional mechanisms, such as delayed identification and removal of corrupted practices, or a generally continuously changing system might render these short term pressures dominant in determining system behavior. However, our model also provides the opportunity to examine the long term system behavior resulting only from effort choice dynamics and cultural evolution (Fig~\ref{fig:Fig4}, 10000 time steps). In a previous similar study, practices were found to inevitable evolve to full corruption \cite{Smaldino2016}. Our model allows to test if a moral incentive component has the power to bound this detrimental evolution. As observed above, the range of existing practices in our model evolves toward the proxy for all levels of competition (Fig~\ref{fig:Fig4}A, D). However, even in the most intense competition, an equilibrium is reached, beyond which the average practice no longer becomes more corrupted (Fig~\ref{fig:Fig4}D). Notably, this equilibrium level of corruption was similar for all levels of competition (Fig~\ref{fig:Fig4}E, middle panel), differing only in the speed with which it was reached (Fig~\ref{fig:Fig4}D). Given the equilibrium corruption level, effort dynamics nevertheless produced an optimal level of competition ($c\approx0.2$). Thus our model demonstrates that an intrinsic motivation to work towards the societal goal can bound the evolution toward fully proxy oriented practices. Note, that the emergent long term equilibrium is analogous to a \textit{lock-in} state, i.e. a stable system orientation toward the proxy, while individual agents i) know the impact of their actions toward and ii) value the societal goal. 

\subsection{Sensitivity/robustness analysis}
Next we tested the sensitivity of these results to parameter variations. Since a full combinatorial exploration of the parameter space is unfeasible, we followed a three-stage strategy: We first systematically varied one parameter at a time holding the other parameters constant (Fig~\ref{fig:Sensitivity}; base value in boldface), monitoring the effect on corruption ($\theta$). We found that parameters fell into three main groups: i) equilibrium determining parameters (Fig~\ref{fig:Sensitivity}A; $ga, gs$), ii) dynamics determining parameters (Fig~\ref{fig:Sensitivity}C, D, $c, sp, m$) and iii) parameters affecting predominantly system variability (Fig~\ref{fig:Sensitivity}B; $t_{sd}, N$) . We then attempted to find deviations from this initial classification by repeating one at a time analysis with a selected range of alternative anchor values,  (Table \ref{tab:Parameter space}, base values in italics).  In over 100 targeted additional model runs, we failed to find such deviations, suggesting our mapping of the main parameter effects was robust. Finally, we undertook two major modification, namely we introduced a step-prospect-function or practice agency, and again found the general relations to hold (with the exception of the system dynamics under practice agency, see below, Fig~\ref{fig:Robustness}).

Across all model specifications equilibrium corruption was determined predominantly by the informational quality of the proxy ($ga$) and the psychological power of the intrinsic incentive towards the societal goal ($gs$; Fig~\ref{fig:Sensitivity}A). In the main model, the dynamics of corruption was predominantly determined by the level of competition ($c$) and the degree to which competition was realized in actual selection events ($sp$; Fig~\ref{fig:Sensitivity}C, D). Practice mutation rate ($m$) had no notable effect on equilibrium or dynamics of corruption for these standard models (Fig~\ref{fig:Sensitivity}D, F, top), but did contribute to the dynamics of corruption when the equilibrium practice was outside the initialization range (Fig~\ref{fig:Sensitivity}D, F, bottom). By contrast, the number of agents in competition ($N$) and talent spread ($tsd$) affected mainly system variability (Fig~\ref{fig:Sensitivity}B). Thus, our model suggests, that equilibrium corruption is determined primarily by proxy information ($ga$) and the intrinsic moral drive toward the societal goal ($gs$). Additionally, the drive toward this equilibrium corruption is governed by the intensity of competition ($c, sp$) and potentially the complexity of the system ($m$). 

\begin{figure}
\centering
\includegraphics[width=\textwidth]{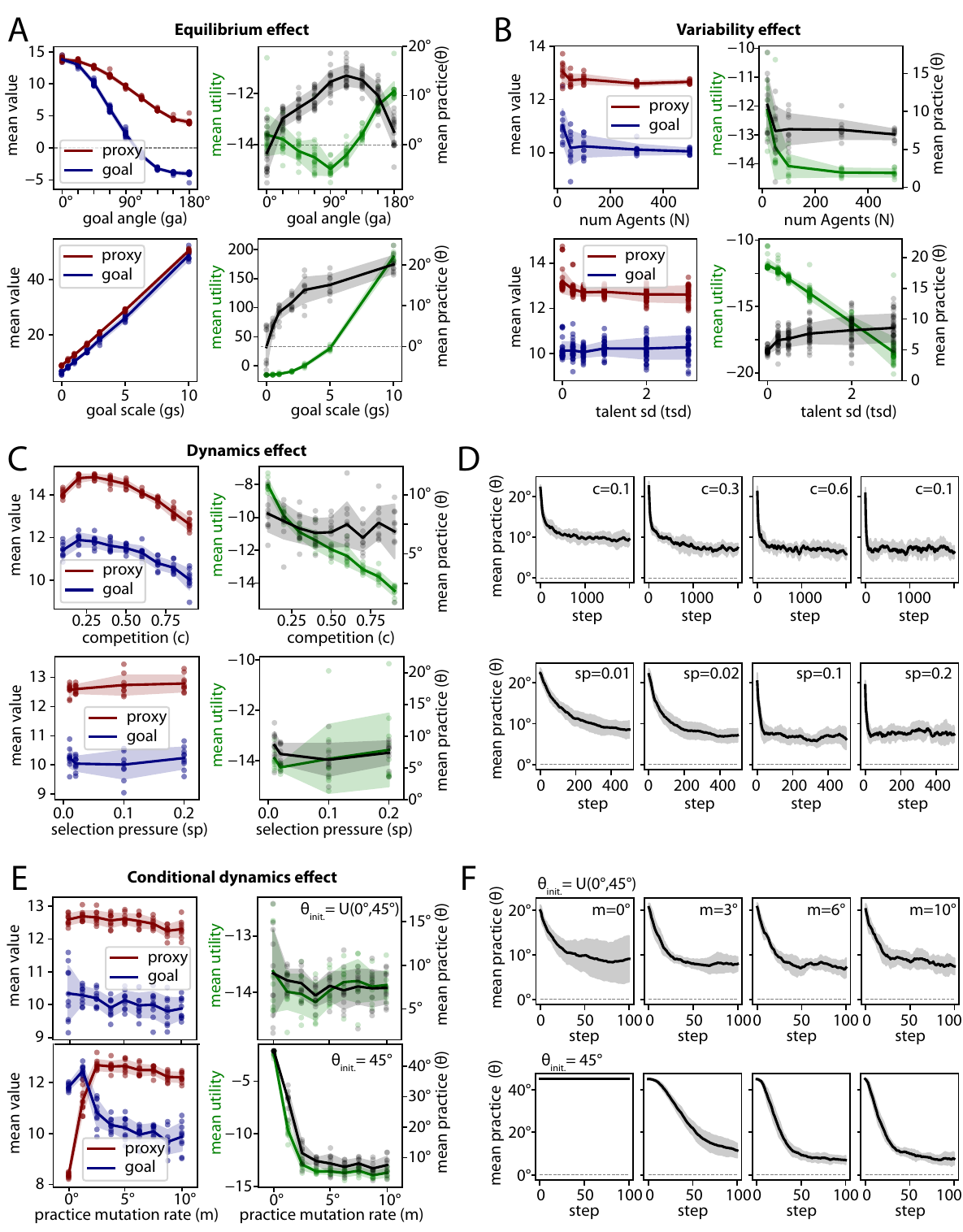}
\caption{\textbf{Sensitivity Analysis}} 
\label{fig:Sensitivity}
\end{figure}

\begin{figure}
\centering
\contcaption{\textbf{Sensitivity Analysis}, parameter base values: $N=100, ga=45^\circ, \theta_{init.}=\mathcal{U}(0^\circ,ga), gs=1, t= \mathcal{N}(10,1), sp=0.1, m=2^\circ$. One parameter at a time was varied while leaving the other parameters at their base values. Overview plots (panels A-C, E) show mean model outcomes at equilibrium $/theta$ (step numbers were increased until convergence to equilibrium was confirmed for each parameter). Data points represent mean population outcomes of individual runs and lines represent $mean \pm sd.$ over runs \textbf{A)} Parameters controlling equilibrium corruption ($\theta$, grey): goal angle ($ga$), top; goal scale ($gs$), bottom. \textbf{B)} Parameters controlling system variability: number of Agents ($N$), top; talent standard deviation ($tsd$), bottom. \textbf{C, D)} Parameters controlling the speed of convergence: competition ($c$), top; selection pressure ($sp$), bottom. ($sp=1$ produced the same equilibrium, but is not shown because the dynamics could no longer be resolved.) \textbf{E, F)} Parameter controlling system dynamics if equilibrium $\theta$ is outside the initialization range. For our standard practice initialization range ($\theta_{init}=U(0^\circ,45^\circ)$) $m$ had no notable effect (top). If all practices were initialized at $\theta_{init}=45^\circ$, $m$ governed the dynamics of convergence, and in the case of $m=0^\circ$ precluded convergence (bottom). } 
\end{figure}

Finally, we tested the robustness of these main findings for two major model modifications, a step-prospect-function and practice agency. First we replaced the prospect function by a simple step function (prospect = -1 for losers, 1 for winners; Fig~\ref{fig:Robustness}A-C). While overall effort and value creation was lower (given the lower maximal prospect differential), the respective effects of the parameters remained robust. This is important, given that the Kahnemann/Tversky type prospect considered primarily may apply to individual agents, but may not hold when larger entities, such as labs or companies, are considered as agents. Second, we introduced agency over the practice angle by letting agents choose not only effort but also practice to maximize utility at every time step. Remarkably, the main determinants of equilibrium corruption, i.e. $ga$ and $gs$, again followed the same pattern (Fig~\ref{fig:Robustness}, D-E). However, introducing practice agency led to a dramatically increased speed of convergence and decreased variability of practices within a population. Indeed, our specification of practice agency was to potent, that it effectively overrode the cultural evolution mechanism and related parameters. Interestingly, it also markedly increased the ability of $gs$ to counteract corruption, leading to approximately twofold higher equilibrium $\theta$ (compare Figs \ref{fig:Sensitivity}A and \ref{fig:Robustness}D). Thus our main finding, that equilibrium corruption is determined by $ga$ and $gs$ was highly robust, while the dynamics determining effects of $c$, $sp$ and $m$ become less relevant when agents continuously choose their practice orientation to maximize utility. These results further suggest, that increased agency over practices may help counteract corruption given sufficient $gs$.

\begin{figure}
\centering
\includegraphics[width=\textwidth]{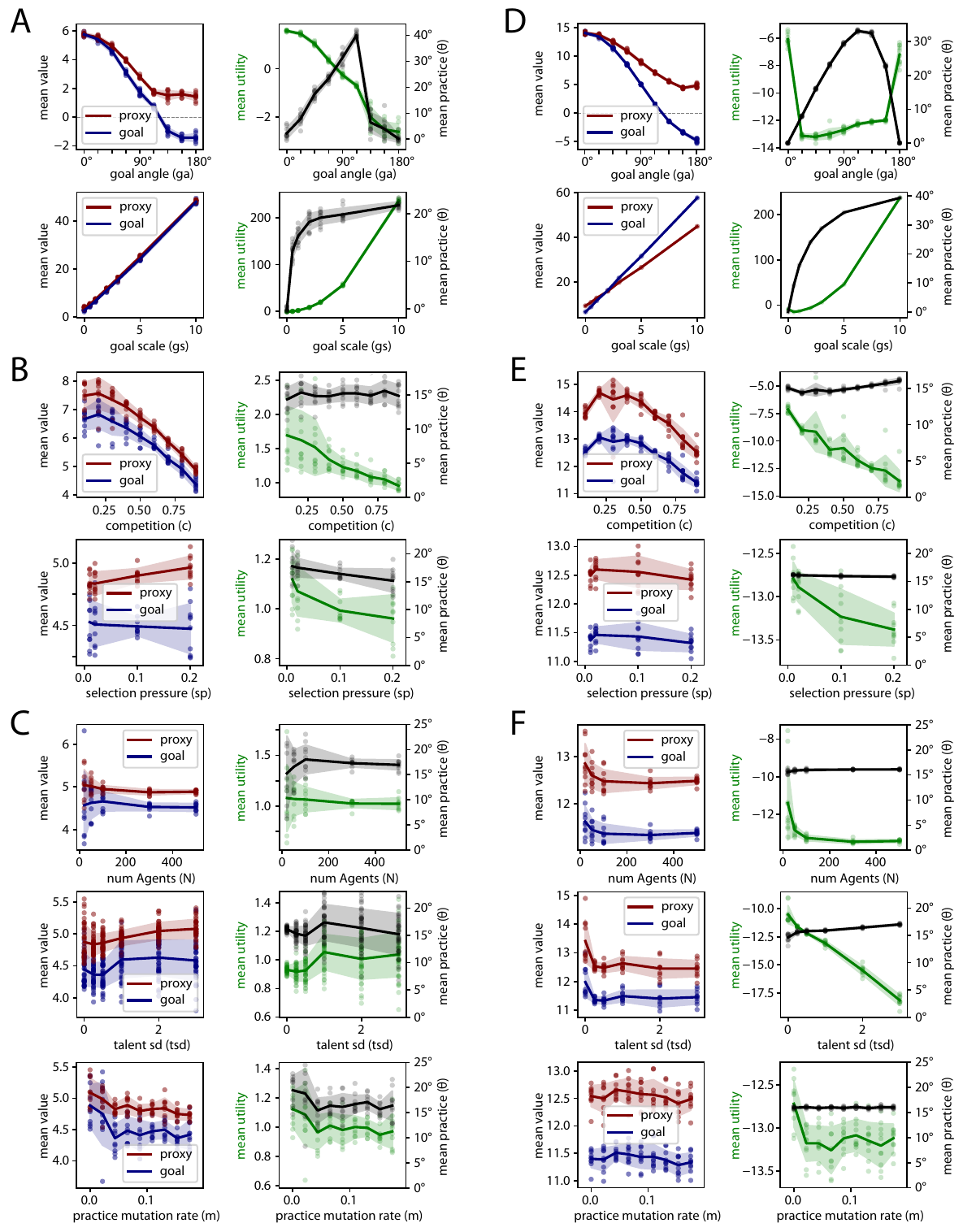}
\caption{\textbf{Robustness}, parameter base values: $N=100, ga=45^\circ, \theta_{init.}=\mathcal{U}(0^\circ,ga), gs=1, t= \mathcal{N}(10,1), sp=0.1, m=2^\circ$. One parameter at a time sensitivity analyses was repeated for two alternative models: One containing a step-function prospect [-1,1]  \textbf{(A-C)} and one containing agency over the practice angle \textbf{(D-F)}} 
\label{fig:Robustness}
\end{figure}

\section{Discussion}
We have presented an agent-based computational model, sketching the central components of a socio-economic theory denoted \textit{Proxyeconomics}. The theory is motivated by the central insight, that any societal competition to achieve an abstract goal must rely on proxy measures. These invariably capture only partial information, and become targets for the competing individuals, resulting in a susceptibility to corruption. 
In the following, we will first discuss the agent-based model (section~\ref{sec:model_disc}) and then proceed to discuss additional implications of proxy-based competition, towards a more general theory of proxyeconomics (section~\ref{sec:theory_disc}).

\subsection{The agent-based model}
\label{sec:model_disc}
In this section we will discuss the more general background of our model, including specific limitations and potential extensions. We will begin by exploring the iterative effort choice algorithm (section \ref{sec:effort_algorithm}), and the cultural evolution mechanism \ref{sec:evolution}). We will then discuss a range of broader informational, psychological and sociological implications (sections \ref{sec:information} - \ref{sec:sociology}), embedding the model in a larger, trans-disciplinary literature. 

\subsubsection{Agent-based contest algorithm}
\label{sec:effort_algorithm}
The relation of our model to the economic contest literature deserves particular mention, since to the best of our knowledge we are the first to implement a neuroeconomically plausible iterative effort choice heuristic in an agent-based model of contests. Our simple, neuroeconomically motivated, contest heuristic contains i) a neural representation of utility, ii) multiple sources of input to this utility and iii) reference dependence of this utility \cite{Glimcher2016}. The sigmoidal prospect function represents a confluence between ecological, psychological and economic theory \cite{Mishra2014}. It requires plausible, noisy input information and computational capacity from individual agents. Intriguingly, the emergent behavior reproduced many empirical findings, the origins of which are still poorly understood \cite{Dechenaux2014}. 
Firstly, effort incentivization is usually optimal at intermediate levels of competition, consistent with experimental and field observations \cite{Harbring2008, Orrison2004, Bradler2016}. Secondly, agents performing far below the survival threshold display a `discouragement effect' \cite{Dechenaux2014}. While traditionally discouragement is thought to result from agent heterogeneity, our model predicts it even in cases without agent heterogeneity, simply due to chance and the emergent distribution of proxy performances. For the same reason, our model predicts a bifurcation of effort, particularly in the presence of heterogeneity, but even without. Additionally, the model makes specific predictions about how competition affects the variance of effort over time and the population. 
At the same time, the model allowed to intuitively capture the signaling/screening effect on talent but also proxy-orientation. This simultaneous screening for talent and proxy-orientation, is suggested to arise in any proxy-based competition under most conditions. Accordingly, observed high performers are likely to generally represent a mix of true high performers (on the goal scale) and highly proxy-oriented agents.
A particularly interesting set of studies in this context are investigations of contests with the possibility of sabotage \cite{Harbring2008, Falk2008}. A mixed strategy containing productive effort as well as sabotage is analogous to a low practice angle in our model. Indeed, increased competition (a smaller fraction of winners or increased prize spread) led to a redistribution of effort towards sabotage, consistent with our results. Notably, studies addressing contest incentives in explicit multitask settings appear to be lacking and thus remain a crucial avenue for future research \cite{Sheremeta2016}. Finally, our iterative effort choice algorithm makes predictions about the temporal evolution of effort expenditure, including its variability in time and the path of convergence to potential equilibria. In this, it resembles previous experimental contest models based on reinforcement learning \cite{vandenBos2013} or belief-updating \cite{Fu2015}. A specific prediction is variability of individual effort over time and over the population, which is proportional to the level of competition (qualitatively matching data presented in \cite{Harbring2008, Dutcher2015}).
Accordingly, we specified what is to our knowledge the first agent-based iterative effort choice algorithm. This generative agent-based algorithm reproduced a remarkable range of empirically observed effort patterns observed in the experimental contest literature, which were not considered during model design \cite{Dechenaux2014, Sheremeta2016}. Embedding this algorithm within a multitasking framework further allowed us to intuitively model the complex screening effect of competition on observed (proxy) as well as unobserved (goal) performance.

\subsubsection{The evolution towards bad practices}
\label{sec:evolution}
A central question we posed of our model, was if such a multitasking mechanism could alter a simultaneously acting mechanism of cultural evolution. Indeed, a previous agent-based evolutionary model of cultural practices had reported a robust evolution toward the worst possible practices \cite{Smaldino2016}. The central unifying element with the present model is the existence of a \textit{practice variable} ($\theta$ in the present study), which alters the relation between true (goal) and selected (proxy) performance, formalizing the assumption that there exist practices which contribute differentially to proxy and goal. Furthermore, this variable can be imperfectly inherited, formalizing the assumption that professional practices are complex and transmitted from senior to junior professionals imperfectly. In our view these assumptions are nearly self evident, and the resulting detrimental evolution becomes a powerful prediction. However, even the authors note, that this prediction seems overly pessimistic and invoke intrinsic incentives as a counteracting power \cite{Smaldino2016}.
Here we have modeled such an intrinsic incentive, and observe that it is able to bound the evolution to fully corrupted practices. Specifically, our model led to an equilibrium corruption level, which was substantial, but still markedly different from full proxy orientation. This result is intuitively appealing and, in fact, better matches the results of a meta-analysis of sample sizes within the same study \cite{Smaldino2016}. The mean proxy-orientation as equilibrium was determined primarily by the informational quality of the proxy ($ga$) and the relative psychological strength of an intrinsic incentive toward the societal goal ($gs$). How both parameters may be determined, or altered, will be further discussed below (sections \ref{sec:information} - \ref{sec:sociology}).
Notably, our model suggests that increased agency over the practice orientation has the ability to further decrease the level of corruption. Such agency may result from training in good research practices, allowing a conscious deviation from questionable norms within a research field.
In the absence of practice agency, the dynamics of corruption in our model were governed by competition ($c$ and  $sp$) and  potentially complexity ($m$). Adding practice agency let the model converge to equilibrium nearly instantaneously, effectively occluding the dynamic effect of $c, sp$ and  $m$. However, in our implementation, practice agency is substantial (with the similar speed and range as effort agency). In real systems practices may be determined by a mix of social norms and individual choice, suggesting outcomes may lie between our models with and without practice agency. The interaction between the evolutionary process and an active practice choice could be investigated in more detail by for instance probing models where practice agency is allowed only stochastically. Indeed, exploratory analysis in such a model suggests, that in such cases the level of competition can mediate between the equilibria with and without practice agency.
Accordingly, our model shows a reliable convergence to a corruption equilibrium. This provides a mechanistic explanation for what we have called \textit{lock-in}. The model further suggests two general strategies to modify the equilibrium, namely i) improving proxy information and ii) promoting goal valuation (for instance through narratives \cite{Benabou2018}, see below). Finally, it suggests competition and agency as central variables determining the dynamics of the system.

\subsubsection{Informational perspective - relation to artificial intelligence/ machine learning}
\label{sec:information}
The general approach of the current model was to view competitive societal systems as information processing devices, which collect information to create the proxy and can ultimately be characterized by their realized competitive decisions. This approach embeds the theory into a larger framework of optimization in complex systems discussed in an extensive literature on complex systems, artificial intelligence and machine learning \cite{Wilson2016, Conant1970, Amodei2016, Manheim2018A, Manheim2018B}. The proxy measure is analogous to an objective function and the competitive pressure corresponds to optimization pressure. We suggest that all the principal types of challenges that occur during machine learning optimization also apply in competitive societal systems, including reward hacking, negative side effects and scalable oversight \cite{Amodei2016}. For instance, the process of Campbell's Law is closely related to reward hacking. Negative side effects include reduced value in dimensions which do not directly impact either proxy or goal.  For instance, \cite{Fochler2016} suggest young scientists progressively constrict their valuation repertoire to competitivity, at the cost of `sociability'. 
Scalable oversight describes the process of balancing the cost of creating the proxy with the benefits from better proxy information. One main reason why proxy measures will imperfectly reflect the societal goal is because they need to be cost efficient. For instance, the information contained in an impact factor could be arguably arbitrarily increased by adding more reviewers or having experiments reproduced. Clearly, the cost of improving the proxy needs to be balanced with the potential detrimental effects of an imperfect proxy. Note that both the cost of improving the proxy and the potential detrimental effects of an imperfect proxy are likely to interact with competition. More frequent competitive evaluations may be more costly to perform while simultaneously incurring a higher unobserved cost due to corruption pressures \cite{Gross2019}. Introducing stochastic and/or delayed assessment and correction mechanisms for corrupted practices may lead to new dynamic equilibira (but see \cite{Smaldino2016}).\\
Importantly, considering which types of information are likely to be lost due to the idiosyncrasies of specific competitive societal systems offers a prime path to predicting corruption. For instance if proxies are assessed in predefined short time intervals, then any outcome with longer time frames is subject to corruptive pressure. Other important considerations include reliance on representatives and sampling in space and time. For instance, market mechanisms collect information in a direct, decentralized and continuous manner while other systems rely on representatives and predefined sampling procedures (e.g. peer reviewers/ political representatives). Finally, if multiple interlocked competitive systems interact it is important to consider if they share or reciprocally counteract informational deficits (see section \ref{markets}). 

Accordingly, the analogy to machine learning suggests three central conclusions: 1. There is an optimal optimization-pressure which crucially depends on the information captured in the proxy. 2. Problems are likely to arise through both continuous system change and the optimization pressure itself, necessitating a continuous higher level assessment process. 3. Analysis of the information generating mechanism of the proxy is likely to provide detailed predictions on patterns of corruption.
In other words, we should continuously expect, attempt to measure, and mitigate corruption in any competitive societal system. Importantly, the term corruption must be understood as an inevitable system-level force arising through actions based on imperfect information, which may, but need not, involve intention.

\subsubsection{Decision theory}
\label{sec:dec_theory}
Our model draws on a large experimental literature addressing the individual level decision mechanism, by including a moral component into decisions. While the adopted economic multitask model provides an elegant way to formalize this, it must be stated clearly, that the actual mechanisms are far from understood. For instance, it is unknown what the relative motivational power of moral and competitive incentives is, particularly for real professionals. Compared to laboratory settings both the moral incentive (e.g. treating a patient well) and the competitive incentive (actual professional survival), may be substantially more powerful.  Nevertheless, experimental investigations \cite{Hochman2016,Pittarello2015,Mazar2008} consistently demonstrate that both moral and egoistic incentives play a role, and provide essential insight about the potential psychological mechanisms during incentive conflicts. Recently, Chugh and Kern \cite{Chugh2016} have even suggested that the need to maintain an ethical self-image is the dominant principle, and that self-interested actions are only permitted to the degree, that this self-image can be maintained. In light of this it is important to note that implications of a decision for the societal goal are likely to be associated with higher ambiguity, longer time frames and less personal relevance than implications for the proxy \cite{Moore2004}. This results directly from the concept of proxy-based competition, since proxy is almost by definition an attempt to make an abstract societal goal concrete, immediate and personally relevant. Of course this has unavoidable consequences for decisions, given the well known phenomena of ambiguity discounting, temporal discounting and social discounting \cite{Jones2009, Weber2008, Strombach2015}.
Another important question, addressed in a separate literature, concerns the effect of competition on decisions \cite{Sheremeta2016}. While a large body of experimental evidence demonstrates the power of competition as an incentive, the underlying mechanisms remain poorly understood, particularly in multitask settings. However, neuroeconomic research is beginning to reveal the important role of loss-aversion in contests \cite{Delgado2008}. More generally, the fields of behavioral economics, psychology and neuroeconomics are beginning to converge towards set of actual, empirically validated, decision mechanisms. Specifically, they suggest a computationally bounded mechanism with multiple (moral/ egoistic), potentially reference dependent, valuation inputs converging into a single utility computation \cite{Glimcher2016, Fehr2011}. While we have attempted to capture these confluent insights into our simple decision model, future research, and more complex empirically validated decision models will unquestionably yield a superior basis for generative agent-based models.\\
Finally, there is substantial empirical evidence for interindividual differences in moral/egoistic drive. For instance, there is direct experimental evidence for an increased propensity to sabotage in males than females \cite{Dato2014}, which may partially explain observed productivity differences concerning the proxy \cite{Lariviere2013}. Variable incentive strengths could be easily modeled as variations in goal scale (or prospect function variability). Incorporating such variability could inform on the outcomes of empirically observed differences between genders or in `machiavellanism' scales \cite{Niederle2011,Tijdink2016}. 
Accordingly, our simple formalization of an agent-decision mechanism is i) neuroeconomically plausible and ii) captures many important empirical findings. However, a large range of additional complexities could be integrated into the model, in order to investigate their implications. The most fruitful way to pursue this would, in our view, be through close interaction between modeling and experimental approaches. 

\subsubsection{Sociological/ Cultural evolution perspective}
\label{sec:sociology}
The sociological/ cultural perspective explicitly acknowledges the complexity of cultural practices and the degree to which these are determined beyond the individual level. We have drawn from the model by \cite{Smaldino2016}, in order to capture the slow evolution of a large body of cultural information implicit in professional practices. Additional, factors which may prove highly relevant to long and short term outcomes, are the network structure between agents and the precise formulation of information transmission between them. In our model, all agents observe the noisy proxy performances of all other agents, implying full network connectivity. Though this may seem unrealistic, the idea that professionals are generally aware of their approximate competitive standing seems justified. To verify robustness against the full connectivity assumption, we probed if restricting the sampled proxy performances to 9 or 22 neighboring agents altered outcomes, but this had no effects other than increasing variability (not shown). Future models could further explore system dynamics if practice angles are influenced by social forces such as the formation of social norms \cite{Epstein2001}. While, social-norm transmission is implied in the hereditary transmission of practices modeled here, they may also be determined more directly through neighboring agents (see e.g. \cite{Benabou2018}). This is likely to introduce additional nonlinear effects in space and time, as locally normative practices emerge or collapse. Notably, similar mechanisms could directly impact the knowledge about, and valuation of, the societal goal ($gs$). For instance, Benabou et. al., \cite{Benabou2018} have recently modeled the spread of moral narratives through populations. Such mechanisms could be included to endogenize $gs$ as a locally determined variable for each agent.
Another partially sociological factor, that has here been simplified into a single exogenous variable is the goal angle, i.e the informational relation between proxy and goal. In real systems, the proxy will be determined by the institutions creating it. From the complex adaptive systems perspective, such institutions are best viewed as autopoietic systems whose existence and stability is not necessarily related to any \textit{societal goal} \cite{Luhman1995, Wilson2016}. A variation of this view is that competitive institutions are primarily self-reinforcing power-structures, and that the \textit{societal goal} may serve only as their public justification. We would argue that the public justification of the system should be viewed as its societal goal. Moreover, the autopoiesis of institutions is likely a necessary requirement for their existence and hence also for any contribution toward the societal goal. Modeling societal institutions as autopoietic entities and probing how they may be brought to additionally serve some abstract societal goal, could help to understand how the goal angle may actually arise.
Accordingly, our simple formalization of cultural evolution captures some important sociological mechanisms, but is dramatically simplified. Nevertheless, we believe that the progressive cultural evolution toward the proxy, described here, should be a central component of any theory of proxy-based societal competition.

In sum, our agent-based model captures a range of crucial implications of proxy-based societal competition. It provides a first attempt to integrate two of the central arising forces, and it's central mechanisms are amenable to experimental verification. Finally, it provides a versatile framework to explore additional complexities, ranging from agent-level decision biases to network effects or implications of agent heterogeneity.

\subsection{Beyond the model}
\label{sec:theory_disc}
In this section we will discuss some broader implications of our theory. We will first outline some striking policy and moral implications of the central premise. Next, we will outline how proxy-based competitions may interlock and act over different scales, e.g. in markets. A final brief section will suggest how we can systematically assess the corruption of real systems, proposing that findings from behavioral economics can be leveraged into predictions of corrupt patterns.

\subsubsection{Policy implications}
Interestingly, a number of familiar policy arguments can be directly derived from the setup of proxy-based competition. For instance, one party may focus on the divergence between proxy and societal goal, advocating costly improvement of proxies or regulation of specific proxy-oriented practices. An opposing party may suggest that regulation or improving proxies will not be cost effective, and that unilateral reorientation toward an ambiguous societal goal will result in i) losing competition and ii) no change in proxy orientation at the system level.
The present theory implies that both arguments result from a deep intuitive understanding of the system and that a confrontation of both views is likely necessary for an optimal regulation of that system. Moreover, it suggests, that the optimal degree and design of regulation will depend on the specifics of a system, where importantly the degree of corruption, the potential cost/benefit of improving the proxy, and the necessity for coordinated rather than unilateral action, can be assessed systematically and scientifically (see e.g. section \ref{sec:information}). One specific, perhaps somewhat counter-intuitive policy prescription to mitigate corruptive pressures, while avoiding associated costs are partial lotteries. Intriguingly, these have been suggested for several domains independently such as politics and science \cite{Pluchino2011, Smaldino2019}.
Finally, it is important to again note that the institutions which create the proxy should at least partially be viewed as self-reinforcing power structures (section \ref{sec:sociology}). This is simply because they are likely to be primarily designed by current proxy winners (professionals which have excelled within the current system). Accordingly, we might generally expect some inertia in societal systems where current proxy winners would decrease their competitive advantage (or the valuation of their life legacy) by questioning the proxy.

\subsubsection{Moral implications}
Another simple yet profound implication of the central setup of proxy-based competition concerns the moral structure of decision problems. So far we have considered only an egoistic incentive, and an incentive toward an abstract societal goal. However, individual competitive success is often linked to others, e.g. one's family, team or employees. In such cases the `egoistic' action may become a moral imperative. For instance, a lead investigator may have to weigh a questionable research practice against the social responsibility of securing funding for her employees. Note, that due to the inherent ambiguity and temporal gradients between proxy and goal, this will tend to imply weighing a relatively certain social harm against a relatively uncertain societal harm. This arguably has direct moral implications. 
Similarly, a CEO may have to weigh her responsibility toward her employees against ambiguous environmental or social damage. Indeed, in practice, securing jobs is frequently invoked as a justification for morally problematic practices, such as selling arms, damaging the environment, or economizing on worker safety and well-being. Such cases demonstrate that this is not a theoretical argument, but a central part of current public moral discourse. The present theory implies that this type of moral dilemma will tend to be automatically created by proxy-based competition. 

\subsubsection{Proxyeconomics across scales - considering markets} \label{markets}
So far we have referred to the institutions/mechanisms which create the proxy as single coherent entities. However, this was an operational choice. In practice these institutions/mechanisms are often complex and may contain proxy-based competition themselves. For instance, the allocation of publication space in high impact journals is determined through competition between authors but also between journals. Such nested, or interlocked proxy-based competitions may counteract or amplify corruption, depending in part on the overlap between informational deficits of the respective proxies. Generally, increasing the scale and complexity of a system is likely to increase the number of failure modes and thereby the scope for corruption \cite{Manheim2018B}. However, the market mechanism arguably represents a system of interlocked competitions which powerfully prevents many kinds of corruption. For instance, in firms, competition for consumers counterbalances competition for capital leading to a product price which incorporates substantial information about consumer valuation as well as production cost. Furthermore, the distributed, continuous, and direct (incentive compatible) processes of markets are likely to prevent many sources of information loss. For instance, markets collect information directly and continuously from consumers rather than from representatives in preset intervals. Nevertheless, above we have suggested whole market economies might become proxy-oriented, implying some degree of conservation or amplification of corruption. Among industrialized nations, this suggestion is supported by macro-level similarities of the observed phenomena, namely the fact that exponential economic growth leads to a statistically negligible changes in subjective as well as objective measures of well-being (e.g. \cite{Kubiszewski2013, Diener2018, Kahneman2010, Stiglitz2010, Wilkinson2009}). This is consistent with data showing that within countries subjective well-being saturates (and sometimes reverses) with increasing income (e.g. \cite{Jebb2018, Diener2018}). However, the self-referentiality of proxy performance in competition suggests that economic growth (proxy-performance) will be pursued regardless. An evolutionary/neuroscientific understanding of human decision making suggests how evolutionary mismatch, status seeking behaviors, and/or preference learning dynamics could be exploited to maintain economic growth without contributing to human well-being \cite{Burnham2016}. In the following, we will go through a concrete example of how evolutionary mismatch and preference learning could be exploited. The example should furthermore demonstrate how the superior efficiency of market mechanisms may amplify rather than counteract corruption across scales. 
Assume consumers display an excessive preference for high sugar products due to evolutionary mismatch. Excessive here means, that if they were to make a fully informed decision, including all aspects of health and long term well-being, they would choose a lower level of consumption. Producers may gain an advantage in competition for consumers by increasing sugar content and obscuring information about negative long term health consequences. Investors may gain a competitive advantage by investing in producers performing these two practices, and thus generating higher profits. If we accept the societal goal of the food-industry, including stakeholders at all levels, as maximizing consumer welfare, then this implies that corruption at one level entails promoting corruption at other levels. For instance companies at all levels have an incentive to obscure detrimental health consequences of high sugar products. This is an example of synergistic corruption paths for proxy-competitions at multiple scales. Similar perverse scenarios are plausible for other addictive products (eg. opioids, gambling, etc). We suspect that obscuring or removing information about detrimental effects of proxy orientation of a shared proxy represents a frequent basis for amplifying corruption between interacting proxy-competitions. For instance, a number of industries maintain an elaborate and costly network of institutions designed to obscure information about detrimental long term health and environmental consequences of their products \cite{Conway2011, Farrell2019}. Indeed, it appears this network has been instrumental in preventing action against global warming.
In this context, we want to note that the theory presented here is closely related to public goods dilemmas, where the socially optimal action (no defection) corresponds to goal orientation and corruption corresponds to coordination failure \cite{Ostrom1999, Manheim2018B}. Indeed, the payoff matrix in a public goods (or prisoners) dilemma can be disaggregated into a \textit{goal} and a \textit{proxy} component, showing how the typical payoff patterns may naturally arise in proxy based competition. The implication is that any proxy-based competition is likely to create a public goods dilemma to some degree (again depending on the information captured in the proxy).

\subsubsection{Applied proxyeconomics or behavioral economics 2.0}
Finally, we want to suggest how the present theory may guide diagnosis and intervention for real societal systems (applied proxyeconomics). A general strategy would be to systematically investigate informational idiosynchrasies of proxy-generation to predict patterns of corruption. Such predictions would allow to diagnose corruption within societal systems, as well as to design mitigation strategies. For instance, we have recently leveraged the scientific phenomenon of positve-publication bias to make a range of otherwise baffling predictions concerning observed scientific sample sizes \cite{Braganza2019}. Positive publication bias here is an informational idiosynchracy of the proxy (the publication record of an author), in that it determines which information is captured and which is lost. Including this idiosyncrasy into a simple model then allowed to make predictions about patterns of sample size choices, which could then be compared to competing goal-orientation accounts as well as empirical data. It has further, recently been used to explore mitigation strategies \cite{Campbell2019}.
Similarly, we believe decision biases identified in behavioral economics can be systmatically leveraged to predict patterns of corruption within societal systems. For instance, excessive proxy-orientation would be expected to prominently shape consumer choice architectures such that they capitalize on known decision biases to increase profit \cite{Hollands2017, Thaler2014, Gabaix2005}. Indeed, behavioral economists emphasize that there is no neutral choice architecture, and suggest that we should consciously and transparently structure choice architectures to systematicaly nudge citizens into beneficial behavior \cite{Thaler2014}. Critics argue that this approach is paternalistic and should be avoided. The present theory suggests, that in the absence of conscious and transparent structuring of choice architectures, \textit{proxy-measures} will be the principal forces. Accordingly, behavioral economics can be used to derive fine grained, falsifiable predictions of actual corruption patterns. Prevalent marketing and advertising practices could be readily analyzed with respect to their implications for proxy- and goal performance, given known decision biases. For instance, advertisements have recently been used to analyze behavioral market failure in the payday lending market \cite{Hawkins2016}. Behavioral economics can then be further used to design choice architectures, which mitigate decision biases and therewith corruption. 
 
\subsection{Conclusion}
We have outlined a transdisciplinary theory of proxyeconomics, which applies whenever a societal system employs proxy measures to mediate competition towards an abstract goal. Our agent-based computational model synthesizes several major insights across disciplines into a formal framework, suggesting a central role of competition on effort expenditure, individual utility, selection and cultural evolution. Accordingly, there may be an optimal level of competition, depending on the complexity of the system and the preferences of the participating agents. Furthermore, we have demonstrated, that an individual-level decision mechanism, which includes an intrinsic goal-oriented motivation component, can bound the evolution to corrupt practices. More generally, the theory provides a conceptual and predictive framework to empirically assess the degree to which actual societal systems may be wastefully or detrimentally oriented towards proxy measures. Importantly, it includes a mechanistic account of how a system can remain \textit{locked in} to a relatively proxy oriented state, even if all individual agents know of, and value, the actual societal goal. This may help to explain and address diverse phenomena such as the scientific reproducibility crisis or inaction to the threat of global warming. 

\section*{Acknowledgements}
I thank Heinz Beck for continuous support. I further thank Christina Selenz, Everard Braganza, Johnathan and Laura Ewell, Klaus G. Troitzsch and Gerben Ter Riet and the participants of various conferences for many helpful comments. The project was funded through the VW-Foundation program \textit{Originalitaetsverdacht}.

\nolinenumbers

\section*{Supporting information}

\paragraph*{S1 Code}
\label{S1_Code}
{\bf Main model code}, Python3 code, based on mesa framework.

\paragraph*{S2 Code}
\label{S2_Code}
{\bf Batch run \& figure generation code: Competition}, Code to run a family of models, analyze outputs and create figures, with competiton as the main input parameter of interest.

\paragraph*{S3 Code}
\label{S3_Code}
{\bf Batch run \& figure generation code: Goal angle}, Code to run a family of models, analyze outputs and create figures, with goal angle as the main input parameter of interest.

\paragraph*{S4 Code}
\label{S4_Code}
{\bf Batch run \& figure generation code: Goal scale}, Code to run a family of models, analyze outputs and create figures, with goal scale as the main input parameter of interest.

\paragraph*{S5 Code}
\label{S5_Code}
{\bf Batch run \& figure generation code: Population size}, Code to run a family of models, analyze outputs and create figures, with the number of agents in the population as the main input parameter of interest.

\paragraph*{S6 Code}
\label{S6_Code}
{\bf Batch run \& figure generation code: practice mutation amplitude}, Code to run a family of models, analyze outputs and create figures, with practice mutation amplitude as the main input parameter of interest.

\paragraph*{S7 Code}
\label{S7_Code}
{\bf Batch run \& figure generation code: Selection pressure}, Code to run a family of models, analyze outputs and create figures, with selection pressure as the main input parameter of interest.

\paragraph*{S8 Code}
\label{S8_Code}
{\bf Batch run \& figure generation code: talent variability}, Code to run a family of models, analyze outputs and create figures, with talent standard deviation as the main input parameter of interest.


\end{document}